\documentstyle[12pt,epsfig]{article}                                                    

\newlength{\dinwidth}                                                    
\newlength{\dinmargin}                                                    
\def\lapproxeq{\lower .7ex\hbox{$\;\stackrel{\textstyle                                                    
<}{\sim}\;$}}                                                    
\def\gapproxeq{\lower .7ex\hbox{$\;\stackrel{\textstyle                                                    
>}{\sim}\;$}}                                                    
\def\be{\begin{equation}}                                                    
\def\ee{\end{equation}}                                                    
\def\bea{\begin{eqnarray}}                                                    
\def\eea{\end{eqnarray}}                                                    
\def\GeV{\rm GeV}
                                                    
\def\funp{{I\!\!P}}                    
                                                    
\begin{document}                                                    
\titlepage                                                    
\begin{flushright}                                                    
IPPP/08/07   \\
DCPT/08/14 \\                                                    
4 March 2008 \\                                                    
\end{flushright}                                                    
                                                    
\vspace*{2cm}                                                    
                                                    
\begin{center}                                                    
{\Large \bf Early LHC measurements to check predictions for central exclusive production}                                                    
                                                    
\vspace*{1cm}                                                    
V.A. Khoze$^{a,b}$, A.D. Martin$^a$ and M.G. Ryskin$^{a,b}$ \\                                                    
                                                   
\vspace*{0.5cm}                                                    
$^a$ Institute for Particle Physics Phenomenology, University of Durham, Durham, DH1 3LE \\                                                   
$^b$ Petersburg Nuclear Physics Institute, Gatchina, St.~Petersburg, 188300, Russia            
\end{center}                                                    
                                                    
\vspace*{2cm}                                                    
                                                    
\begin{abstract}                                                    
We show how the early data runs of the LHC can provide valuable checks of the different components of the formalism used to predict the cross sections of central exclusive processes. The `soft' rapidity gap survival factor can be studied in electroweak processes, such as $W$+gaps events, where the bare amplitude is well known. The generalized gluon distribution, in the appropriate kinematic region, can be probed by exclusive $\Upsilon$ production. The perturbative QCD effects, especially the Sudakov-like factor, can be probed by exclusive two- and three-jet production. We discuss the possible role of enhanced absorptive corrections which would violate the soft-hard factorization implied in the usual formalism, and suggest ways that the LHC may explore their presence.  
\end{abstract}                                          
     
\section{Introduction}

Central exclusive production is now recognized as an important
search scenario for new physics at the LHC, see for instance, \cite {KMRProsp}~-~\cite{cr} and references therein.
The experimental studies of such processes are at the heart
of the FP420
project \cite{LOI,cox} which proposes to complement the
CMS and ATLAS experiments at the LHC
by installing  additional
forward proton detectors 420m away from the interaction region.
In particular, these detectors will allow the measurement of the exclusive
production of new heavy particles, such as Higgs bosons.
As demonstrated in \cite{KKMRext}~-~\cite{fghpp} such measurements will be able
to provide valuable information on the Higgs sector of MSSM and other popular
BSM scenarios.

Indeed, central exclusive processes are very interesting both from the viewpoint of theory, since they contain a mixture of soft and hard QCD effects, and of experiment, as they provide a clean environment to measure the quantum numbers and masses of new objects which may be seen at the LHC. Moreover, the $J_z=0$ selection rule ($J_z$ is the
projection of the total angular momentum along the proton
beam direction), arising in central exclusive diffractive processes \cite{KMRmm, DKMOR},
provides a unique possibility to study directly the coupling of
the Higgs-like bosons to the bottom quarks, because the LO QCD
background is strongly suppressed by the $J_z=0$ rule. As is well known,
the determination of the $Hb\bar {b}$ Yukawa coupling
appears to be very difficult for other search channels at the LHC.

\begin{figure}
\begin{center}
\includegraphics[height=6cm]{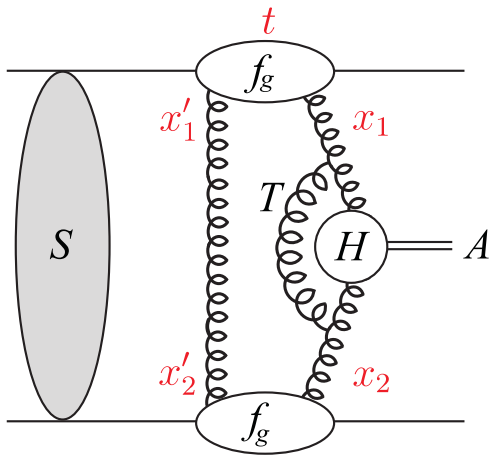}
\caption{A symbolic diagram for the central exclusive production of a system $A$.}
\label{fig:parts}
\end{center}
\end{figure}

The theoretical formalism \cite{KMR,KMRProsp,KMRsoft} needed to describe a central exclusive diffractive process of a system $A$ contains quite distinct parts, shown symbolically in Fig.~\ref{fig:parts}. In brief, we first have to calculate the $gg \to A$ hard subprocess, $H$, convoluted with the gluon distributions $f_g$. Next we must account for the higher loop corrections which reflect the absence of additional QCD radiation in the hard subprocess -- that is, for the Sudakov suppression factor $T$. Finally we must enter soft physics to calculate the survival probability $S^2$ of the rapidity gaps either side of $A$ -- that is the probability that the exclusive nature of the process will not be destroyed by the secondaries produced by the rescattering of the incoming particles.

\section{Objective}
The uncertainties associated with the prediction of the rate of a central exclusive process are potentially not small. Each of the above stages has its own uncertainties. Therefore, it is important to perform checks of the approach using processes with appreciable cross sections that will be experimentally accessible in the first data runs of the LHC with integrated luminosities in the range 100 ${\rm pb}^{-1}$ to 1 ${\rm fb}^{-1}$. Here, our aim is to identify processes where the different ingredients of the formalism used to calculate central exclusive production can be tested experimentally, more or less independently.

The outgoing protons in the forward regions of the main LHC detectors
will be measured by the proton tagging detectors (Roman Pots). At 220 m on either side of the CMS detector there exists the roman pots of
the TOTEM experiment \cite{TOTEM,CMS-TOTEM}.  ATLAS will have the ALFA detector \cite{ALFA} at 240m as well as the proposed RP220 detector \cite{RP220} at 220m.
As we already mentioned, FP420 project \cite{LOI,cox}
proposes to install forward proton taggers for both ATLAS and CMS. Note that RP220 and FP420 aim to operate
at high luminosities.
However it is quite likely that for the first 1-2 years of LHC running
the forward proton detectors will not be operational. We therefore first consider measurements which do not rely on tagging the forward outgoing protons.

Even without proton tagging, diffractive measurements at the LHC
can be performed through the detection of rapidity gaps.
This is a well known technique used extensively
at HERA and the Tevatron.
A summary of the forward detectors instrumented around ATLAS and
CMS is given, for instance, in \cite{forw}. The central detectors (CD) of CMS and ATLAS have an acceptance in pseudorapidity
$\eta$ of roughly $|\eta |<2.5$ for tracking information
and $|\eta |<5$ for calorimeter information.  We will discuss a situation where a heavy system $A$ is detected in the region $|\eta |<2.5$, and where the calorimeters in $2<\eta <5$ interval, are used to select events with rapidity gaps. 

In the present paper the word {\it gap} means a rapidity interval $\Delta\eta$ devoid of hadronic activity -- for the charged particles we assume a `track veto', while the absence of neutrals should be
checked by the calorimeters. The amplitudes of all the
processes discussed below are infrared stable and are not affected by the
possibility of soft gluon emission. 

Due to angular ordering, which originates from coherence, in those processes where the gap is provided by $W$-boson exchange (as in Fig.~\ref{fig:WZ}(a,c) below), the only possible soft gluons occur
at the edge of the gap and arise from the corresponding quark jet.
These gluons should be accounted for in the jet searching algorithm. In cases where the gaps are associated with colour-singlet, two-gluon exchange (as in central exclusive dijet production), the presence of Sudakov-like $T$-factors in the unintegrated
gluon distributions (see eq.(\ref{eq:rat6}) below) guarantee that there is no emission of any additional soft gluons.

The selection of rapidity gap events by a `veto' trigger 
can be used up to rather large luminosities, when the mean number, $N$, of interactions per bunch crossing is sizeable. However at larger luminosities
the efficiency of the trigger is reduced by a factor $e^{-N}$ -- that is, by the probability
to have no additional `pile-up' inelastic interaction in the bunch crossing.
This probability can be measured independently in the same experiment.

Of course, the proposal to use calorimeters in the $2<\eta <5$ interval to
select the events with a rapidity gap does not mean that we will only consider gaps with $\Delta \eta<3$. First, part of the gap can be at smaller $\eta$ and, secondly, extensive additions are foreseen to enlarge the coverage in the
forward regions\footnote{CMS will have the CASTOR calorimeter
operating with the coverage of $5.1< |\eta |<6.5$
and a Zero Degree Calorimeter (ZDC) with an acceptance for neutral
particles with $\eta >8$.
 Both calorimeters have an electromagnetic
and hadronic section. Moreover, CMS is expected to have integrated read out with
TOTEM (see \cite{TOTEM}) allowing CMS to benefit from the TOTEM forward
coverage and TOTEM, in turn,  from the CMS central coverage \cite{CMS-TOTEM}.
ATLAS plan a Cerenkov detector LUCID with an acceptance
 $5.4 < \eta < 6.1$ and a ZDC with an acceptance $8.3 < \eta < 9.2$.
Diffractive studies are under discussion also at ALICE, see for instance,
\cite{RS}.
The ALICE detector has a central barrel covering the pseudorapidity range
$-0.9 < \eta < 0.9$ and
(on one side) a
muon spectrometer covering the region of $2.4 < \eta < 4$
and a ZDC. Additional detectors for trigger purposes and for event classification
are placed on both sides of the central barrel, such that the range
$-3.7 < \eta < 5$ is covered. This configuration allows the possibility of a (double)
rapidity gap trigger by requiring no activity in the event classification
detectors, see \cite{RS}.}. Possible ``holes'' in particle observation in small rapidity intervals between different calorimeters do not affect our predictions very much. The probability to produce extra soft hadrons in the hole region only is proportional to the effective Pomeron-Pomeron cross section which is rather small, according to the triple-Regge analysis of UA8 data \cite{UA8}. We evaluate the correction to be less than 3$\%$ for a hole of size $\Delta \eta_{\rm hole} \sim 1.5$.

The main uncertainties of the predictions for exclusive processes are associated with the calculation of
\begin{itemize}
\item [(i)] the probability $S^2$ that additional soft secondaries will not populate the gaps separating the centrally produced system $A$ from the outgoing protons (or the products of their dissociations)\footnote{There are quite a few theoretical studies of the $S^2$ factor,
starting from the publications \cite{DKS,BJ}
to more recent ones \cite{KMRsoft,GLMrev} and references therein.
Following Bjorken (\cite{BJ}) this quantity is often called the survival
probability or soft survival factor.};
\item [(ii)] the probability to find the appropriate gluons that are given by generalized unintegrated distributions $f_g(x,x',Q_t^2)$;
\item [(iii)] the higher order QCD corrections to the hard subprocess amplitude, where the most important is the so-called Sudakov suppression caused (in the Feynman gauge) by the double-log loop, denoted $T$ in Fig.~\ref{fig:parts}, see \cite{KMR};
\item [(iv)] the so-called semi-enhanced absorptive corrections (see, for example, \cite{KKMR,bbkm,KMRln}) and other effects which may violate the soft-hard factorization (which is implied by Fig.~\ref{fig:parts}).
\end{itemize}

We discuss below, in turn, possible checks of the various ingredients of the calculation of the exclusive cross sections. To be precise, we address the uncertainties (i)-(iv) in Sections 3-6 respectively.

\section{Gap survival probability $S^2$}

As a rule, the gap survival probability, that is the probability that the secondaries produced in additional soft interactions do not populate the rapidity gaps, is calculated within a multichannel eikonal approach. This method of calculation implicitly assumes a factorization between the soft and hard parts of the process. The probability of elastic $pp$ rescattering, shown symbolically by $S^2$ in Fig.~\ref{fig:parts}, can be calculated in a model independent way once the elastic $pp$ amplitude is known, that is, the elastic cross section $d\sigma_{\rm el}/dt$ is measured at the LHC. However there may be some excited states (corresponding to $N^*$ resonances and low mass diffractive dissociation) between the blob $S$ and the perturbative QCD amplitude on the right-hand-side of Fig.~\ref{fig:parts}. The presence of such additional states enlarges the absorptive correction. Usually this effect is calculated using a Good-Walker formalism \cite{GW} with a two- or three-channel eikonal. In order to experimentally check the role of this effect, we need to consider a process with a bare cross section that can be reliably calculated theoretically. Good candidates are the production of $W$ or $Z$ bosons with rapidity gaps on either side.

\subsection{$W$ production with rapidity gaps} 
\begin{figure}
\begin{center}
\includegraphics[height=6cm]{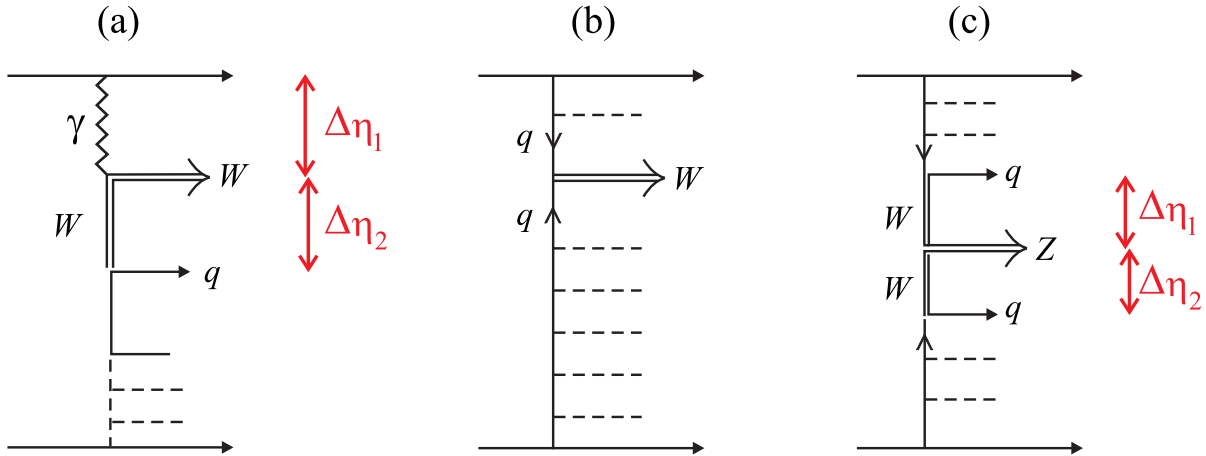}
\caption{Diagrams for (a) $W$ production with 2 rapidity gaps, (b) inclusive $W$ production, and (c) $Z$ production with 2 rapidity gaps.}
\label{fig:WZ}
\end{center}
\end{figure}
In the case of `$W$+gaps' production the main contribution comes from the diagram shown in Fig.~\ref{fig:WZ}(a) \cite{KMRphoton}. One gap, $\Delta \eta_1$, is associated with photon exchange, while the other, $\Delta \eta_2$ , is associated with $W$ exchange. The cross section of this process can be calculated straightforwardly. It is proportional to the quark distribution in the proton at a large scale and not too small $x$, where the uncertainties of the parton densities are small. To select these events at the expected luminosities of the early data runs, we can use the rapidity gap veto trigger combined with a high $p_t$ decay lepton or jet trigger. The momentum transferred through the photon is typically small and the rapidity gap $\Delta \eta_1$ runs from the rapidity of the $W$ boson, $y_W$, up to the maximum rapidity measured. The gap, $\Delta \eta_2$, corresponding to $W$ exchange, is limited by the rapidity of the quark jet, which balances the transverse momentum of the $W$ boson (which has a broad distribution with $q_{tW} \sim M_W/2$ since it is driven by the $t$-channel $W$ propagator).

At first sight, the probability for soft rescattering in such a process is rather small and we would expect $S^2 \sim 1$. The reason is that the transverse momentum, $k_t$, distribution of the exchanged photon is given by the logarithmic integral
\be
\int \frac{dk^2_t~k_t^2}{(|t_{\rm min}|+k_t^2)^2},
\ee
for which the dominant contribution comes from the low $k_t^2$ region. In other words, the main contribution comes from the region of large impact parameters, $b_t$, where the opacity of the proton is small. However the minimum value of $|t|$,
\be
|t_{\rm min}|~\simeq~\frac{m_N^2 \xi^2}{1-\xi}
\label{eq:tmin}
\ee
is not negligibly small. Note that the momentum fraction $x_p=1-\xi$ associated with the upper proton can be measured with sufficient accuracy\footnote{The CDF collaboration \cite{cdf} have demonstrated that this method provides an accurate determination of $\xi$.}, {\it even without the tagging of the forward protons}, by summing the momentum fractions
\be
\xi_i~=~\sqrt{m_i^2+k_{ti}^2}~e^{y_i}/\sqrt{s}
\label{eq:xi}
\ee
of the outgoing $W$ and the hadrons observed in the calorimeters.  As long as the gap $\Delta \eta_2$ is large, the dominant contribution to the sum $\xi=\sum \xi_i$ comes from the decay products of the $W$ boson\footnote{Even for the leptonic decay channels, the momentum of the $W$ boson can be reconstructed with knowledge of the missing transverse energy and $M_W$.}. For example, for $\eta_W=2.3(-2.3)$, we expect an $\xi$ distribution centred about $\xi \sim 0.1(0.001)$.

\begin{figure} 
\begin{center}
\includegraphics[height=9cm]{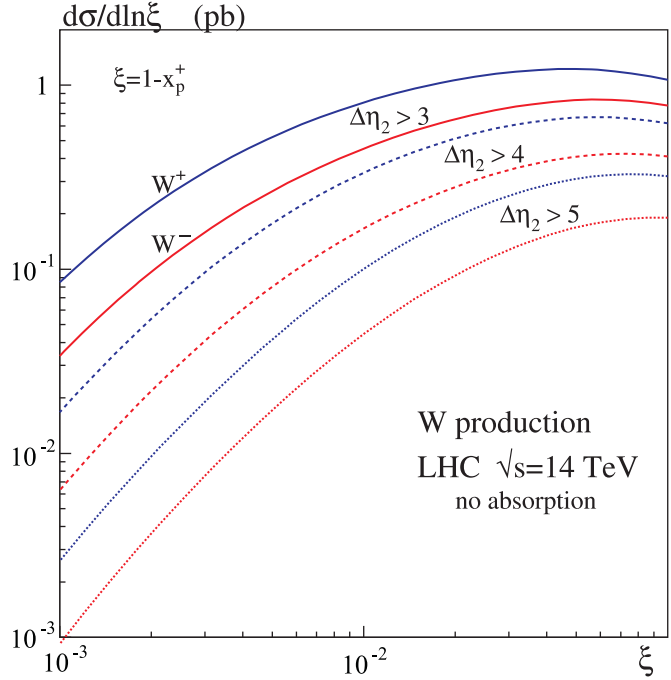}
\caption{The cross sections for $W$+gaps events as a function of $\xi$ at the LHC for different choices of the size of the rapidity gap $\Delta \eta_2$. No suppression from rescattering effects is included. The momentum fraction of the proton that is carried by the photon is $\xi=1-x^+_p$, where $x^+_p$ is the momentum fraction of the outgoing upper proton that emits the photon in Fig.~\ref{fig:WZ}(a). We use the superscript + to indicate that the cross sections correspond to the configuration where the photon is emitted by the proton going in the + direction. The event rate observed in the central detector will be twice as large as that shown.}
\label{fig:W1}
\end{center}
\end{figure}
\begin{figure} 
\begin{center}
\includegraphics[height=9cm]{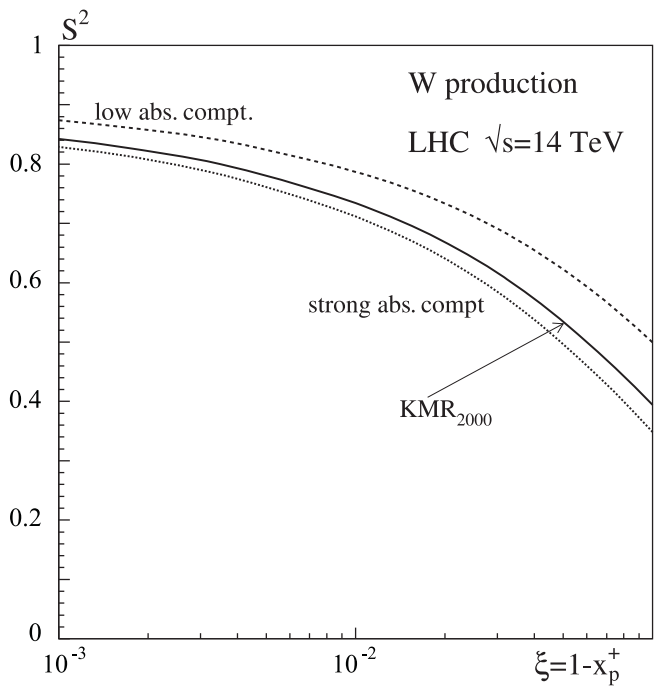}
\caption{The rapidity gap survival factor $S^2$ as a function of $\xi$ calculated using the global soft model of \cite{KMRsoft}. }
\label{fig:W2}
\end{center}
\end{figure} 
\begin{figure} 
\begin{center}
\includegraphics[height=9cm]{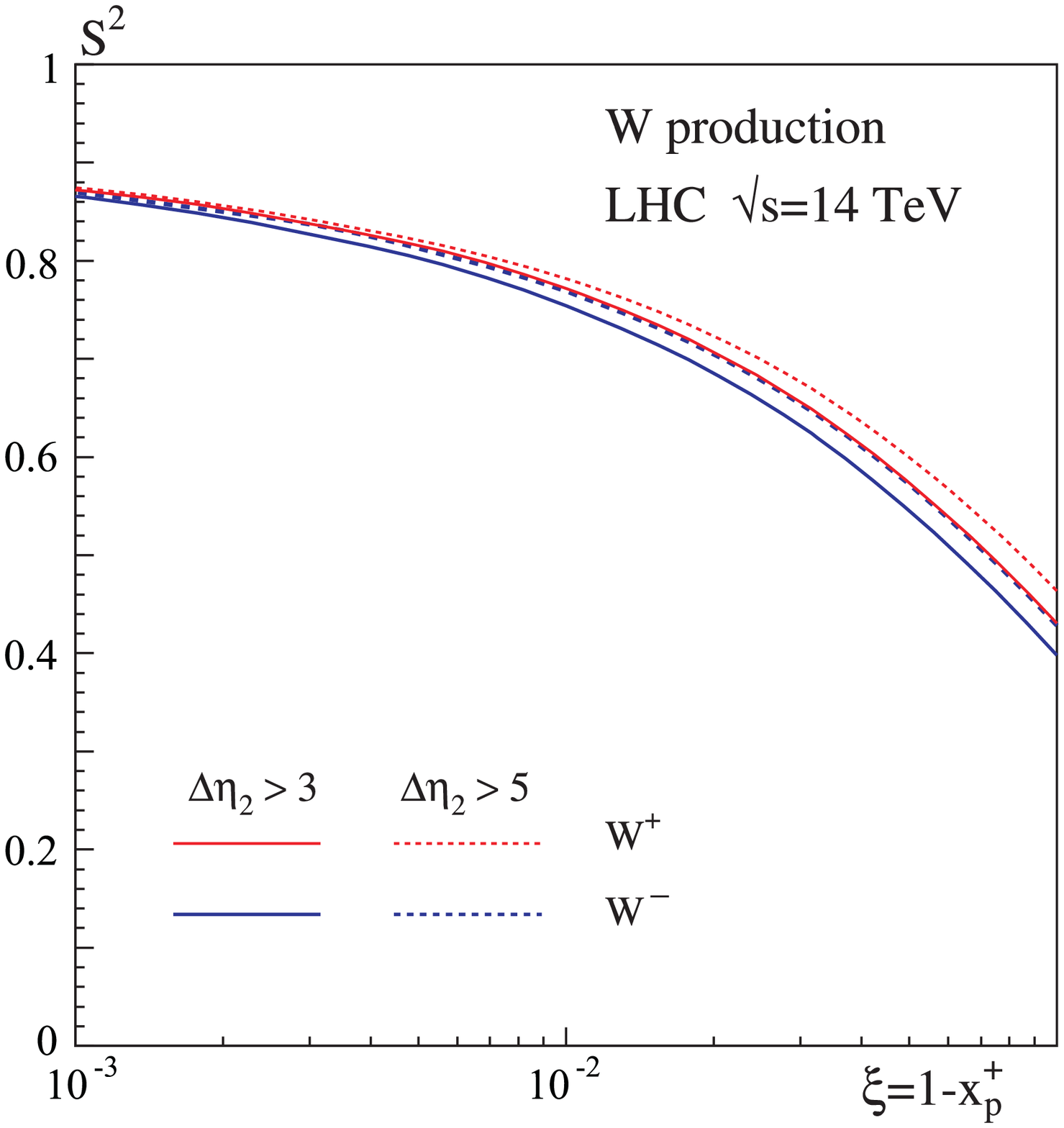}
\caption{The rapidity gap survival factor $S^2$ as a function of $\xi$ calculated using the global soft model of \cite{KMRsoft}, assuming that the valence (sea) quarks are associated with the weak (strong) absorptive components. The small spread of the predictions arising from the different partonic content of the diffractive eigenstates mean that $W$+gaps events offer a meaningful test of the $S^2$ factor. Note that $S^2$ for the $W^+$ signal  is larger since it has a bigger valence quark contribution.}
\label{fig:W3}
\end{center}
\end{figure}
The predictions are shown in Figs.~\ref{fig:W1}--\ref{fig:W3}. First, in Fig.~\ref{fig:W1}, we neglect the absorptive correction, and show the cross section for $W$+gaps events for different choices of the gap size $\Delta \eta_2$. We see cross sections of up to 1 pb, depending on the gap size and on $\xi$. The rescattering decreases the cross section by the factor $S^2$ shown in Figs.~\ref{fig:W2} and \ref{fig:W3}. Already, at this stage, we face the possibility of a violation of soft-hard factorization. We use the Good-Walker formalism \cite{GW} to describe multichannel eikonal rescattering. That is, we use the diffractive eigenstates $\phi_k$ which diagonalize the `nucleon'-Pomeron couplings $\beta_{ij}$, which describe the transition from nucleon excited state $i$ to state $j$. In Ref.~\cite{KMRsoft} it was assumed that each eigenstate has the same size and the same partonic composition. The continuous curve in Fig.~\ref{fig:W2} was calculated under this assumption. On the other hand, in the extreme cases where we assume that the partons which participate in the process of Fig.~\ref{fig:WZ}(a) are concentrated in the diffractive eigenstate with the lowest (largest) absorptive cross section, we obtain the results shown by the dashed (dotted) curves. A rather more realistic scenario (see, for example, \cite{KKMR}) is to allocate the valence quark to the component with the smallest absorption and the sea quark to the component with largest cross section. The corresponding predictions are shown in Fig.~\ref{fig:W3}. Since the valence quark contribution is more important for $W^+$ production and for the configuration with the largest gap size $\Delta\eta_2$, the expected gap survival factor $S^2$ is found to be larger. Now, the spread of predictions caused by the different partonic content, shown in Fig.~\ref{fig:W3}, is rather small. With a more realistic assumption (in which each diffractive state contains some part of the valence and some part of the sea) the spread will be even smaller.

In the first LHC data runs it may be difficult to measure the absolute value of the cross section with sufficient accuracy. Most probably the ratio ($W$+gaps/$W$ inclusive) will be measured first. In this case, the inclusive $W$ production process (in the same kinematic region, Fig.~\ref{fig:WZ}(b)) plays the role of the luminosity monitor. Note that the cross section for inclusive $W$ production is much larger than that with rapidity gaps. The reason is an inclusive $W$ is produced directly by $q\bar{q}$ fusion, which is prohibited for gap events since the colour flow produced by the $t$-channel quarks populates the $\Delta \eta_{1,2}$ rapidity gaps. 

Of course, the survival factor $S^2$ measured in $W$+gaps events is quite different from that for exclusive Higgs production, which comes from smaller values of $b_t$. Nevertheless this measurement is a useful check of the model for soft rescattering.

\subsection{$Z$ production with rapidity gaps} 

It would appear that a good way to study the low $b_t$ region directly is to observe $Z$ boson production via $WW$ fusion, see Fig.~\ref{fig:WZ}(c). Here, both of the rapidity gaps originate from heavy boson exchange and the corresponding $b_t$ region is similar to that for central exclusive Higgs production\footnote{Even so it may not be exactly the same $b_t$ region, since, in general, the impact parameter distributions of quarks and gluons may be different.}. The expected $Z$+gaps cross section is of the order of 0.2 pb, and $S^2$=0.3 for $\Delta \eta_{1,2} > 3$ and for quark jets with $E_T>50$ GeV \cite{pw}. Again it would be sufficient to measure the ratio ($Z$+gaps/$Z$ inclusive).

One problem is that even with the $E_T>50$ GeV cut, the QCD background arising from the QCD $b\bar{b}$ central exclusive production is comparable to the electroweak $qq \to Z+2$ jet signal. Therefore we should concentrate on the leptonic decay modes of the $Z$ boson, which results in a smaller event rate\footnote{Note that in the recent study \cite{nikit} it was demonstrated that the so-called Track Counting Veto (TCV) is robust for selection of the central
rapidity gap events in vector-boson fusion $H \to \tau^+\tau^-$
searches at CMS.
The idea of track counting is close in spirit to the hadron level
gap selection of \cite{DKS,pw} and, as compared to the (more) standard
calorimeter jet veto technique, the TCV has various advantages. In particular, it
does not involve  calorimeter
scale uncertainties, suffers less from pile-up contributions
and so can be used at higher luminosities.
Moreover, it is planned to apply this technique
at CMS using data on central $Z \to \mu^+\mu^-$ events.
Though the original motivation for this analysis \cite{nikit} is to
study the detector conditions, such a measurement, in particular,
presented in terms of the ratio ($Z$+gaps/$Z$ inclusive) may appear to be
one of the first tests of models for soft survival factor. }.

\subsection{With tagged protons}

So far we have discussed measurements which do not depend on observing the forward protons. However, when the forward proton detectors become operational we can do more. Both the longitudinal and transverse momentum of the forward protons can, in principle, be measured. Hence we can study the $k_t$ behaviour of the cross section section for $W$+gap events, and scan the proton opacity, as described in \cite{KMRphoton}.

\section{Generalized, unintegrated gluon distribution $f_g$}
The cross section for the central exclusive production of a system $A$ is calculated using $Q_t$ factorization, and essentially has the form \cite{KMR}
\begin{equation}
\sigma(pp \to p+A+p) ~\simeq ~\frac{S^2}{b^2} \left|\frac{\pi}{8} \int\frac{dQ^2_t}{Q^4_t}\: f_g(x_1, x_1', Q_t^2, \mu^2)f_g(x_2,x_2',Q_t^2,\mu^2)~ \right| ^2~\hat{\sigma}(gg \to A).
\label{eq:M}
\end{equation}
The first factor, $S^2 $, is the soft survival factor discussed in the previous section and the factor $1/b^2$ arises from the integration over the transverse momentum of the forward proton assuming the form $d\sigma /dp^2_t \propto {\rm exp}(-bp^2_t)$. Also, $f_g$ denotes the generalized, unintegrated gluon distribution in the limit of $p_t \to 0$.  Below we give a more precise definition of $f_g$ and explain how it is determined.

The generalized gluon distribution has not yet been measured explicitly. However, for the case of interest, where the skewedness is small, it can be obtained from the conventional diagonal gluon distribution, $g$, known from the global parton analyses.  The procedure is as follows. We consider the skewedness first in the transverse momenta, and then in the longitudinal momenta, carried by the two $t$-channel gluons in Fig.~\ref{fig:parts}. The transverse momentum, $p_t$, transferred through the rapidity gap in a central exclusive process is limited by the proton form factor. This is not a large transverse momentum: $p_t^2 \simeq -t \ll Q_t^2$. Typically, we have the hierarchy
\be
|t_{\rm min}| \lapproxeq 10^{-4} ~{\rm GeV}^2; ~~~~~~~p_t^2 \sim 0.2 ~{\rm GeV}^2;~~~~~~~Q_t^2 \sim 4~\GeV^2;
\ee
which follows, respectively, from (\ref{eq:tmin}) with the relevant $\xi \sim 0.01$ or less; from the proton form factor; and, finally, from the presence of the Sudakov-like form factor, $T$ of (\ref{eq:T}), in the integrand of (\ref{eq:M}) -- the maximum of the integrand occurs at $Q_t^2 \sim 4~\GeV^2$, see \cite{KMRProsp,KMRmm}.

As a consequence, it is reasonable to assume the following factorization of the gluon distribution: $F(t)~f_g(x,x',Q^2_t,\mu^2)$. That is, we work in terms of the distribution $f_g(x,x',Q^2_t,\mu^2)$, which is skewed only in longitudinal momentum fractions, $x \ne x'$ in Fig.~\ref{fig:parts};
$Q_t$ is essentially the transverse momentum of each $t$-channel gluon. Finally, $\mu^2 \sim M_A^2/4$ is the factorization scale which separates the partonic distribution from the matrix element of the hard subprocess $gg \to A$. The precise scale, to be used, is given in (\ref{eq:Delta}) below. 

In the region of interest
\be
\left(x' \sim \frac{Q_t}{\sqrt{s}}\right)~~~\ll~~~\left(x \sim \frac{M_A}{\sqrt{s}}\right)~~~\ll~~~1.
\ee
The generalized distribution can be obtained using the Shuvaev transform \cite{shuv} which assumes that the Gegenbauer moments of the generalized parton distributions are equal to the Mellin moments of the diagonal distributions. The accuracy of this assumption is $O(x^2)$. As a result we obtain~\cite{MR01}\footnote
{In the actual computations we use a more precise form as given by eq.(26) of Ref.\cite{MR01}}.
\be f_g(x,x',Q_t^2,\mu^2) = R_g\frac{\partial}{\partial \ln Q_t^2} \left(\sqrt{T(Q_t,\mu)}\: xg(x,Q_t^2)\right),
\label{eq:rat6} 
\ee
where $R_g$ accounts for the $x \ne x'$ skewedness. Note that the double log Sudakov suppression $T(Q_t,\mu)$ is now included in the unintegrated gluon distribution $f_g$, since to provide $Q_t$ factorization we choose an axial/planar gauge (and not the Feynman gauge used to draw $T$ in Fig.~\ref{fig:parts}).
The Sudakov form factor $T$ ensures that the gluon remains untouched in the evolution up to
the hard scale $\mu$, so that the rapidity gaps survive. It results from resumming the
virtual contributions in the DGLAP evolution, and is given by \cite{WMR}
\begin{equation}
T(Q_t,\mu) = \exp\left(-\int_{Q_t^2}^{\mu^2}\frac{\alpha_S(k_t^2)}{2\pi}\frac{dk_t^2}{k_t^2}
\int_\Delta^{1-\Delta}\left[zP_{gg}(z) + \sum_q P_{qg}(z)\right]dz\right), \label{eq:T}
\end{equation}
where $\Delta$ is specified in (\ref{eq:Delta}) below.

The main uncertainty here comes from the lack of knowledge of the integrated gluon distribution $g(x,Q_t^2)$ at low $x$ and small scales. For example, taking $Q_t^2=4~\GeV^2$ we find that a variety of recent MRST \cite{mstw} and CTEQ \cite{cteq} global analyses give gluons which have a spread of
\be
xg~=~(3-3.8)~~{\rm for}~~x=10^{-2}~~~~{\rm and}~~~~xg~=~(3.4-4.5)~~{\rm for}~~x=10^{-3}.
\ee
These are big uncertainties bearing in mind that the cross section for central exclusive production depends on $(xg)^4$.

\subsection{Exclusive $\Upsilon$ production as a probe of $f_g$}
\begin{figure} [t]
\begin{center}
\includegraphics[height=5cm]{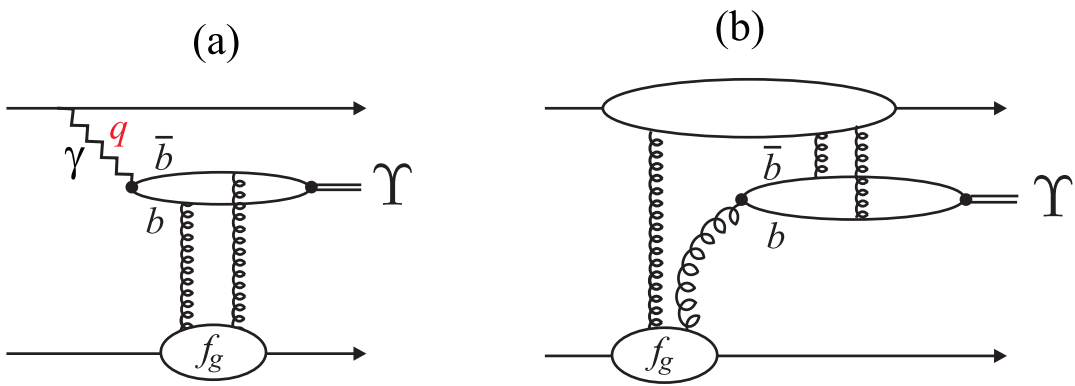}
\caption{Exclusive $\Upsilon$ production via (a) photon exchange, and (b) via odderon exchange.}
\label{fig:upsilon}
\end{center}
\end{figure} 
To reduce the uncertainty associated with $f_g$ we can measure exclusive $\Upsilon$ production. The process is shown in Fig.~\ref{fig:upsilon}(a).  The cross section for $\gamma p \to \Upsilon p$ \cite{mrt} is given in terms of exactly the same generalized, unintegrated gluon distribution $f_g$ that occurs in Fig.~\ref{fig:parts}. If the $\Upsilon$ is detected at the LHC in the rapidity interval $(|y|< 2.5)$ then it will sample the $x_{1,2}$ intervals ($10^{-4}-10^{-2}$). To probe even lower $x$ we can either detect more forward $\Upsilon$ production at ALICE or LHCb \cite{LHCb} or study exclusive $J/\psi$ production, see \cite{KMRphoton}.

Of course, there may be competition between production via photon exchange, Fig.~\ref{fig:upsilon}(a), and via odderon exchange, where the upper three $t$-channel gluons form the C $=-1$ odderon state\footnote{Here the colour indices of the gluons are convoluted with the $d_{abc}$ tensor.}, see Fig.~\ref{fig:upsilon}(b). To date, odderon exchange has not been observed. On the other hand, a lowest order perturbative QCD calculation indicates that the odderon process (b) may be comparable to the photon-initiated process (a), see, for example, \cite{bmsc}. If the upper proton is tagged, it will be straightforward to separate the two mechanisms since odderon production has no $1/q^2$ singularity characteristic of the photon. Photon exchange populates mainly the low $q_t$ region, that is at low $p_t$ of the outgoing proton, whereas odderon production occurs at relatively large $p_t$. Without a proton tagger, an odderon would be revealed by a measured cross section that is larger than that predicted for photon exchange. Exclusive $\Upsilon$ (or $J/\psi$) production could be the first hint of the odderon's existence.

Returning to process (a), we note that the cross section is of the form
\be
\sigma^{(a)}~=~N_\gamma \sigma(\gamma p \to \Upsilon p)
\ee
where the photon flux $N_\gamma$ is well known. For small $q_t$, we may neglect the proton form factor and use the leading log approximation 
\be
N_\gamma~=~\frac{\alpha_{\rm em}}{\pi}\frac{d\xi}{\xi}\frac{ dq^2_t}{q^2_t}.
\ee
The expression for $\sigma(\gamma p \to \Upsilon p) \propto f_g^2$ is given in \cite{mrt}. The cross section \cite{bmsc}
\be
\left.\frac{d\sigma^{(a)}}{dy}\right|_{y=0}~~\simeq~~50~{\rm pb}
\ee
at the LHC energy\footnote{The estimate assumes that the forward protons are not tagged, and so is enhanced by a factor of about 2 caused by the dissociation of the lower proton in Fig.~\ref{fig:upsilon}(a).}. The signal will be diluted by the $\Upsilon \to \mu\mu$ branching fraction of 0.025. 

In order to use this process to constrain the gluon distribution
of the proton it would be preferable to tag the lower proton.
Otherwise there will be some admixture of proton excitations and to calculate precisely the gap survival factor $S^2$,
 as before, we will have to resolve the partonic content of the different
diffractive eigenstates. On the other hand, for low values of $\xi$,
the expected value $S^2$ is close to 1 (see the continuous curve in Fig.~\ref{fig:W2}),
and it may be sufficient to use the existing
HERA data for the ratio of the cross sections of diffractive
$J/\psi$ photoproduction with and without the proton dissociation,
to extract the contribution to the cross section from 
`elastic' events in which the lower proton does not dissociate.

\section{Improvements to the hard subprocess}

The hard matrix elements necessary to predict the cross section for a central exclusive process are calculated using perturbative QCD mainly at leading log accuracy \cite{KMRProsp}. Of course, there exist numerous NLO results, but as a rule these are applicable to inclusive processes which do not allow for the fact that the centrally produced system $A$ must be in a colour singlet state and obey the $J_z=0$ selection rule, that is the projection of the total angular momentum along the incoming proton directions should be zero. Moreover the relation between the generalized, unintegrated gluon distribution, $f_g$, and the conventional diagonal gluon, $g$, was also based on a LO calculation \cite{MR01}. 

Numerically, the largest effects may come from the next-to-leading-log (NLL) corrections to the double log Sudakov-like factor $T$ of (\ref{eq:T}). To include these NLL corrections phenomenologically we choose the limits of the integration over $k^2_t$ in such a way to reproduce the result of an explicit first-loop calculation. It has been shown (see footnote 9 in \cite{KKMRext}) that this is achieved by choosing 
\be
\Delta~=~\frac{k_t}{\mu +k_t}~~~~{\rm with}~~~~\mu=0.62M_A
\label{eq:Delta}
\ee
in (\ref{eq:T}). 

On the other hand, the lower limit $Q_t^2$ of the integration has not yet been validated at this level. Clearly the contribution from the very small $k_t$ region vanishes due to destructive interference between the emissions from the active gluon ($x$) and the screening gluon ($x'$). However, variation of the lower limit from $Q^2_t/2$ to $2Q^2_t$ may alter the prediction by up to an order of magnitude.

Fortunately, the contribution from the region of $k_t \sim Q_t$ can be calculated with better precision. We are seeking the next-to-leading log correction to the double logarithmic $T$ factor of (\ref{eq:T}), that is, for the single log terms in $T$. Therefore, in the region $k_t \sim Q_t$ we need to keep only the LO BFKL-like term which contains the longitudinal logs. This term is given by the usual BFKL kernel \cite{bfkl} which sums all the leading $\alpha_s$log$M_A$ contributions. If we consider the $k_t^2$ integration in (\ref{eq:T}) with a lower cut-off of $k_0^2$, then the above summation amounts to the replacement
\be
\int_{k_0^2}\frac{d^2k_t}{k^2_t}... ~~~~\to~~~~\int_{k_0^2}\frac{d^2k_t}{k^2_t}\left(1-\frac{Q^2_t}{k^2_t+(\vec{Q}_t-\vec{k}_t)^2}\right)...~~~~=~~~~\int_{Q_t^2}\frac{d^2k_t}{k^2_t}... ~.
\label{eq:TTT}
\ee
First, we notice in the middle expression for the integral above that for $k_t \gg Q_t$ the last term in the brackets is negligible, while for $k_t \ll Q_t$ the whole expression in brackets goes to zero and removes the infrared divergence. Now we can put the infrared cut-off $k_0=0$. The result of an explicit calculation of the middle expression for the integral shows that it is equivalent to the integral in (\ref{eq:T}) with the lower limit given by $Q^2_t$, as indicated in (\ref{eq:TTT}).

Thus, in summary, both the upper and lower limits of the $k_t^2$ integral in (\ref{eq:T}) are fixed so as to reproduce the one-loop contributions. The potentially large ambiguity in the value of $T$ has been removed.

To check experimentally the formalism used in the perturbative QCD calculations for the central exclusive matrix element we should study an exclusive process with the emission of one additional jet. Note that the physical origin of the Sudakov $T$ factor is that it expresses the probability ${\it not}$ to emit additional gluons. The formula for $T(Q_t,\mu)$ of (\ref{eq:T}) can be written as $T={\rm exp}(-n)$ where $n$ is the mean number of gluons emitted in the interval $(Q_t,\mu)$. Thus, the observation of the explicit emission of additional gluon jets would provide a direct test of the formalism. Since the central system must be colour neutral, we need {\it either} consider the emission of two extra jets {\it or} to have the possibility to compensate the colour of one emitted gluon by rearrangement of the colour content of the system $A$. The optimum choice is to observe the emission of a third jet in the production of a pair of high $E_T$ jets, that is, in high $E_T$ dijet production.

\subsection{Three-jet events as a probe of the Sudakov factor}

Traditionally, the search for the exclusive dijet signal at the Tevatron, $p\bar{p} \to p+jj+\bar{p}$, is performed \cite {CDFjj} by plotting the cross section in terms of the variable
\be
R_{jj}~=~M_{jj}/M_A~.
\label{eq:jj}
\ee
Ideally, we might expect exclusive dijet production to show up as a narrow peak centred at $R_{jj}=1$, since, for these events, the mass of the dijet system, $M_{jj}$, is equal to the mass, $M_A$, of the whole central system.  Unfortunately, in practice, the $R_{jj}$  distribution is strongly smeared out by
QCD bremsstrahlung, hadronization, the jet searching algorithm and other experimental effects \cite {CDFjj,KMRrj}.

\begin{figure}
\begin{center}
\includegraphics[height=7cm]{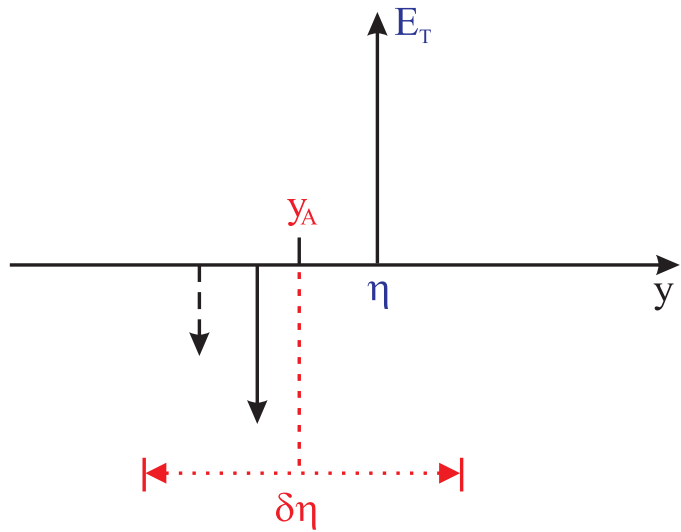}
\caption{The rapidities of the three jets in the central system. Note that the rapidity $y_A$ of the whole central system does not necessarily occur at $y=0$. The rapidity
interval containing the three jets is denoted by $\delta\eta$, outside of which there is no hadronic
activity.\label{fig:2}}
\end{center}
\end{figure}
To weaken the role of this smearing it was proposed in Ref.~\cite{KMRrj} that the observed dijet distribution be studied in terms of a new variable
\be
R_j~=~2E_T ~({\rm cosh}~\eta^*)/M_A~,
\label{eq:j}
\ee
where only the transverse energy $E_T$ and the rapidity $\eta$ of the jet with
the {\it largest} $E_T$ are used in
the numerator.  Here $\eta^* = \eta -y_A$ where $y_A$ is the rapidity of the whole central system\footnote{Note
that the transverse momentum of the dijet system can be neglected, since it is
very small compared to the $E_T$ resolution.}.  Clearly
the jet with the largest $E_T$ is less affected by hadronization, final parton radiation etc.  In particular, final state radiation at lowest order in $\alpha_S$ will not affect $R_j$ at all, since it does not change the kinematics of the highest $E_T$ jet used to evaluate (\ref{eq:j}). Even with the emission of an extra jet during the final parton shower, we will still have $R_j=1$. Thus, to see the role of QCD radiation on the $R_j$ distribution, we only have to account explicitly for additional gluon radiation in the initial state.  At leading order, it is sufficient to consider
the emission of a third gluon jet, as shown in Fig.~\ref{fig:2}, where we take all three jets to lie in a specified rapidity interval $\delta\eta$.  The reason why it is sufficient to consider only one extra jet, is that the effect of the other jets, which, at LO, carry lower energy due to the strong ordering, is almost negligible in terms of the $R_j$ distribution.
\begin{figure} 
\begin{center}
\includegraphics[height=12cm]{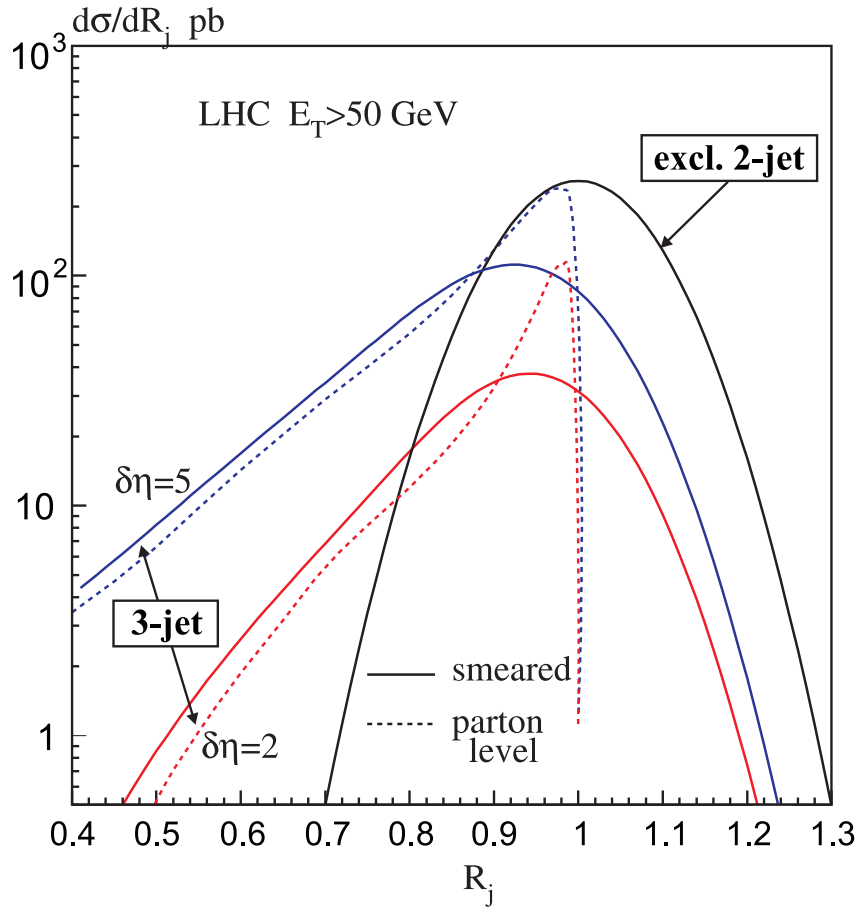}
\caption{The $R_j$ distribution of exclusive two- and three-jet production at the LHC.
Without smearing, exclusive two-jet production would be just a $\delta$-function at $R_j=1$.
The distribution for three-jet production is shown for two choices of the rapidity interval, $\delta\eta$, containing the jets; these distributions are shown with and without smearing. Here, we have taken the highest $E_T$ jet to have $E_T>50$ GeV. To indicate the effect of jet smearing, we have assumed a Gaussian distribution with a typical
resolution $\sigma=0.6/\sqrt{E_T~{\rm in~ GeV}}$. }
\label{fig:Rj}
\end{center}
\end{figure} 

The cross section $d\sigma/dR_j$, as a function of $R_j$, for the exclusive production of a high $E_T$ dijet system accompanied by a third (lower $E_T$) jet was calculated and discussed in detail in \cite{KMRrj}.  The mass of the central system, $M_A$, can be accurately measured by the missing mass reconstructed from the tagged protons, or, failing that, by accurately summing both the light-cone momentum fractions $\xi^+$ and $\xi^-$ of the hadrons observed in the calorimeter, see (\ref{eq:xi}).  As shown in \cite{clp}, the usage of the $R_j$ variable
is quite beneficial for the extraction of exclusive dijet events.
In particular, the dependence on the jet selection criteria (for example, the cone radius and $R$-parameter) is less marked for the $R_j$
than for the conventional $R_{jj}$ variable.  Further discussion can be found in \cite{KMRrj}. Moreover, note that  studying the $R_j$ distribution, we do not need to select events with a rather large $E_T$ of the third jet. To calculate $R_j$ it is sufficient to measure only the $E_T$ of the largest $E_T$ jet, together with the rapidities of first and second jets, and the values of $M_A$ and $y_A$ as measured via
$\xi^+$ and $\xi^-$.

In Fig.~\ref{fig:Rj} we show the $R_j$ distribution of both exclusive two- and three-jet production expected at the LHC. For three-jet production we show predictions for two choices of the rapidity interval $\delta\eta$ within which the third jet must lie. If we take the largest $E_T$ jet to have $E_T>50$ GeV at the LHC, we see that the cross section for exclusive three-jet production reaches a value of the order of 100 pb. Of course, if we enlarge the rapidity interval $\delta\eta$ where we allow emission of the third jet, then $d\sigma/dR_j$ will increase, see Fig.~\ref{fig:Rj}. Indeed, the measurement of the exclusive two- and three-jet cross sections {\it as a function of $E_T$} of the highest jet allow a check of the Sudakov factor; with much more information coming from the observation of the $\delta\eta$ dependence of three-jet production.  Note that the background from double-Pomeron-exchange should be small for $R_j \gapproxeq 0.5$, and can be removed entirely by imposing an $E_T$ cut on the third jet, say $E_T>5$ GeV.

Another way to observe the effect of the Sudakov suppression is just to study the $E_T$ dependence of exclusive dijet production. On dimensional grounds we would expect $d\sigma/dE_T^2 \propto 1/E_T^4$. This behaviour is modified by the anomalous dimension of the gluon and by a stronger Sudakov suppression with increasing $E_T$. Already the existing CDF exclusive dijet data \cite{CDFjj} exclude predictions which omit the Sudakov effect. In the $E_T$ interval from 10 to 35 GeV the Tevatron observations show that the cross section falls by an order of magnitude faster than the prediction \cite{pes}, based on \cite{pes2}, which does not include the Sudakov suppression, see Fig. 20a of \cite{CDFjj}.

It is clear that precise measurements of the $E_T$ behaviour of the exclusive dijet cross section at the LHC offer the possibility to probe the corrections to (\ref{eq:T}). However the study of exclusive three-jet events will provide much more information and, indeed, allow a study of the integrand of (\ref{eq:T}).

\section{Soft-hard factorization: enhanced absorptive effects}
\begin{figure} 
\begin{center}
\includegraphics[height=6cm]{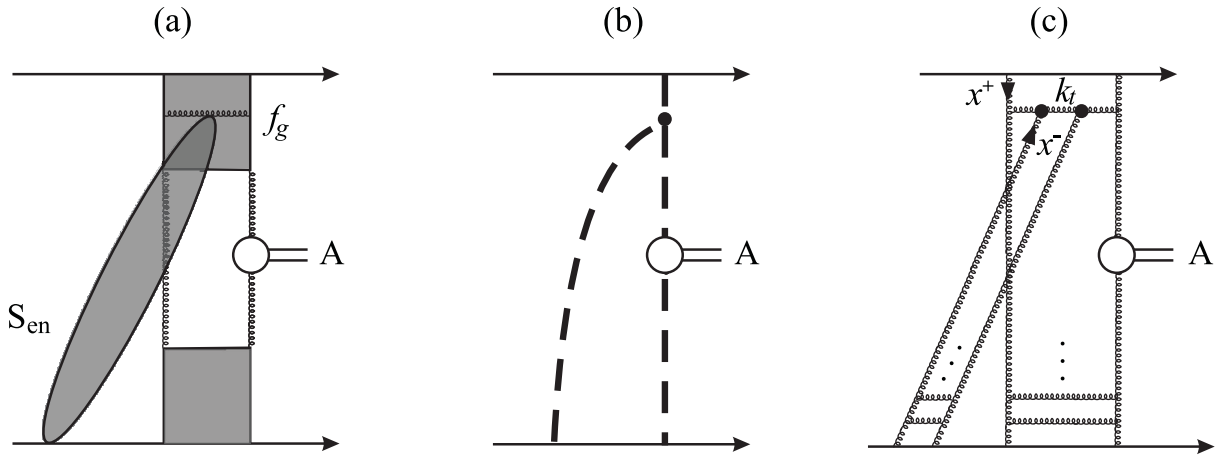}
\caption{(a) A typical enhanced diagram, where the shaded boxes symbolically denote $f_g$, and the soft rescattering is on an intermediate parton, giving rise to a gap survival factor $S_{\rm en}$; (b) and (c) are the Reggeon and QCD representations, respectively.}
\label{fig:enh}
\end{center}
\end{figure}

We mentioned in Section 3 that the soft-hard factorization implied by Fig.~\ref{fig:parts} may be already violated if the different diffractive eigenstates of the $pp$ interaction have different partonic distributions. To date, there is no unambiguous model to distribute the partons obtained in the global analyses between the different diffractive components. However, this is not a major uncertainty.  It is seen in Fig.~\ref{fig:W3} that the difference between $W^+$ and $W^-$, which have a different valence quark contribution, is not large. 

Another potential source of the violation of soft-hard factorization arises from the so-called enhanced Reggeon diagrams, which occur from the rescattering of an intermediate parton generated in the evolution of $f_g$. Such a diagram is shown in Fig.~\ref{fig:enh}(a). Fig.~\ref{fig:enh}(b) is the Pomeron representation of the diagram, where the rapidity of the intermediate parton fixes the position of the triple-Pomeron vertex. Clearly such diagrams will violate the soft-hard factorization.

The contribution of the first Pomeron loop diagram, Fig.~\ref{fig:enh}(b) was calculated in perturbative QCD in Ref.~\cite{bbkm}. A typical perturbative diagram is shown in Fig.~\ref{fig:enh}(c). For LHC energies it was found that the probability of such rescattering may be numerically large\footnote{Of course, the higher-order Reggeon diagrams will considerably reduce the size of the effect. Moreover, allowing for the more complicated multi-Pomeron vertices makes the suppression of the first Pomeron loop contribution even stronger. This effect has be seen in a particular model, see Fig. 17 of \cite{KMRnewsoft}.
It is practically impossible to quantify the reduction within the framework of perturbative QCD.}. The reason is that the gluon density grows in the low $x$ region and, for low $k_t$ partons, approaches the saturation limit. The recent estimates of \cite{WZ} indicate that in the region of $x^- \sim 10^{-6}$ (which is relevant for the LHC kinematics) the value of the saturation scale, $Q_S$, exceeds 1 GeV. In other words, a parton with $k_t<Q_S$ will receive an important absorptive correction. However, this is true for the central region of impact parameter space. On the other hand, eikonal absorption, which is shown in Fig.~\ref{fig:parts} and discussed in Section 3, makes the centre of the disk almost black. This eikonal factor, $S^2$, already forces the central exclusive signal to occur at relatively large $b_t$, namely $b_t>0.5-0.6$ fm, see Fig. 22 of \cite{KMRnewsoft}. In this peripheral region of the proton, the value of $Q_S$ is rather low. In fact, $Q_S^2 \sim 0.3~ \GeV^2$ is found in Fig. 11 of \cite{WZ}. As a consequence, the enhanced diagram will affect only the very beginning of the QCD evolution -- the region that cannot be described perturbatively and which, in our calculation of the cross sections for central exclusive processes, is already included phenomenologically -- that is, described in terms of the multichannel eikonal framework.

This is probably one reason why semi-enhanced screening corrections are not seen in leading neutron production measured at HERA \cite{HERAln}. As the energy of the virtual photon increases, the available rapidity region grows, so the number of intermediate partons that may participate in the rescattering increases, and we expect that the fraction of leading neutrons should decrease. This is not seen in the data \cite{HERAln}. Further discussion is given in \cite{KKMR,KMRln,KMRneutr}.

\subsection{Pomeron-$p$ interactions as a probe of the enhanced effect}

Experimentally, it appears at first sight, that we may study the role of semi-enhanced absorption by observing the $W$+2 gaps process shown in Fig.~\ref{fig:WZ}(a). To do this one may vary the transverse momentum of the accompanying quark jet, $q$, and the size of the rapidity gap $\Delta \eta_2$. For lower transverse momentum of the quark jet we expect a stronger absorptive effect, that should decrease with increasing $\Delta \eta_2$, since the number of partons in the rest of the rapidity interval increases. Unfortunately, the photon-initiated process of Fig.~\ref{fig:WZ}(a) occurs at large impact parameter $b_t$ where the probability of rescattering is small. Moreover it will be very difficult to observe the quark jet at low $E_T$ in the main calorimeter $(-5<\eta<-3)$. Therefore we will discuss other processes which depend on rescattering on intermediate partons. 

\begin{figure} 
\begin{center}
\includegraphics[height=7cm]{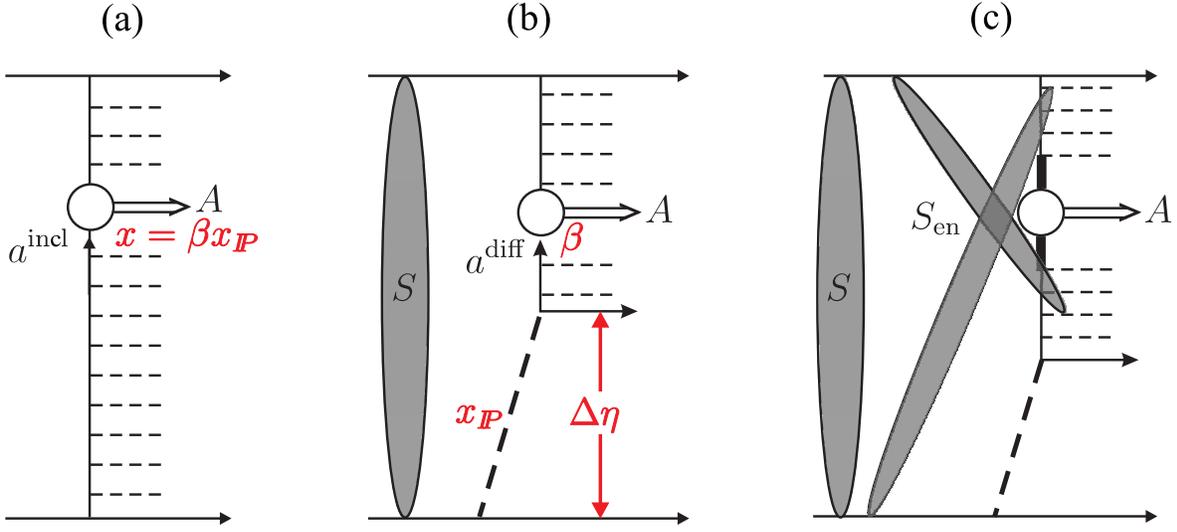}
\caption{Schematic diagrams for (a) the inclusive production of a system $A$, (b) and (c) for the diffractive production of $A$ without and with `enhanced' soft rescattering on intermediate partons. The system $A$ is taken to be either a $W$ boson or an $\Upsilon$ or a pair of high $E_T$ jets.}
\label{fig:3en}
\end{center}
\end{figure}
The observations we have in mind are the measurements of ratio $R$ of diffractive (one-gap) events for $W$ (or $\Upsilon$ or dijet) production as compared to the number of events for the inclusive process (shown in Fig.~\ref{fig:WZ}(b) for $W$ production). These processes are shown schematically in Fig.~\ref{fig:3en}. In other words, $R$ is the ratio of the process in diagram (c) to that in diagram (a). That is
\be 
R~~=~~\frac{{\rm no.~of}~ (A+{\rm gap)~ events}}{{\rm no.~of~ (inclusive}~A)~{\rm events}}~~=~~
\frac{a^{\rm diff}(x_\funp ,\beta,\mu^2)}{a^{\rm incl}(x=\beta x_\funp,\mu^2)}~\langle S^2S^2_{\rm en}\rangle_{{\rm over}~b_t},
\label{eq:Ren}
\ee
where $a^{\rm incl}$ and $a^{\rm diff}$ are the parton densities determined from the global analyses of inclusive and diffractive deep inelastic scattering data, respectively. The heavy central system $A$ is either $W$ or a pair of high $E_T$ jets or $\Upsilon$ or a Drell-Yan $\mu^+\mu^-$ pair. For $W$ or $\mu^+\mu^-$ pair production the parton densities $a$ are quark distributions, whereas for dijet or $\Upsilon$ production they are mainly gluon densities. The diffractive parton densities are known from analyses \cite{mrw2,h1} of diffractive deep inelastic data. Thus measurements of the ratio $R$ will probe the gap survival factor averaged over the impact parameter $b_t$.

If we neglect the effect of enhanced rescattering on intermediate partons, that is set $S_{\rm en}^2=1$ then $R$ is the ratio of process (b) to (a). However this ratio will be reduced by the rescattering on the intermediate partons as sketched in Fig.~\ref{fig:3en}(c)). The largest contribution to the extra suppression factor $S_{\rm en}^2$ comes from the interaction of the upper intermediate partons with the lower proton, since the partial energy corresponding to this interaction is larger, which leads to a larger absorptive cross section. Note that the enhanced effect is practically forbidden in a rapidity interval close to the rapidity of $A$ (shown by the bold lines in Fig.~\ref{fig:3en}(c)), as the transverse momenta of partons in this region are quite large. The mean number of partons, $\langle N \rangle$, emitted in the evolution from 1 GeV up to the hard scale $\mu \sim M_A/2$ is given by the power of the exponent in the $T$ factor of (\ref{eq:T}). We have $\langle N \rangle \sim 1$ or 2 for $M_A \simeq$ 10 or 100 GeV respectively, that is for exclusive $\Upsilon$ or $W$ production. These are the partons with $p_t>Q_S(b_t)$ which do not suffer the enhanced absorptive effect. As a rule, each parton occupies about one unit of rapidity. Therefore in the evaluation of the enhanced effect we exclude from our calculation an appropriate rapidity interval on either side of $A$, which is shown by the bold lines in Fig.~\ref{fig:3en}(c). The presence of a ``threshold'' factor, that is the existence of a rapidity interval on either side of $A$ where the enhanced contribution is effectively forbidden, was emphasized in \cite{KKMR}. An analogous threshold strongly suppresses the enhanced absorption to exclusive Higgs production \cite{KMRln}.

In the region of relatively large $b_t$ the value of $Q_S$ is low. Hence we evaluate the enhanced absorption in terms of Reggeon framework, using the phenomenological triple-Pomeron coupling extracted from diffractive $J/\psi$ production \cite{KMRj}. The absorptive corrections for $J/\psi$ production are low, and the value $g_{3\funp} \sim g_N/3$ should be close to the original bare triple-Pomeron coupling. Here, $g_N$ is the coupling of the Pomeron to the proton. The contribution of the one-loop Regge diagram of Fig.~\ref{fig:enh}(b) can be quite large. Indeed, its strength, as compared to the original cross section,
can be estimated by evaluating Fig.~\ref{fig:enh}(b). We obtain
\be
r~=~\int \frac{4(g_{3\funp}/g_N)\sigma_{pp}(\eta')}{16\pi^2B}~d\eta',
\ee
where $\sigma_{pp}(\eta')$  is the proton-proton cross section evaluated
at the energy $\sqrt{s}=m_N\exp(\eta'/2)$.  The factor 4 comes from the AGK cutting rules \cite{AGK} and $B \simeq 5~ \GeV^{-2}$ is the sum of the $t$ slopes of the form factors involved. The integral is taken over the appropriate rapidity intervals of the intermediate partons in Fig.~\ref{fig:3en}(c). Since the result is sizeable, we allow for the summation of the higher order diagrams by replacing the absorptive factor $1-r$ by $S^2_{\rm en}={\rm exp}(-r)$.

\begin{figure} 
\begin{center}
\includegraphics[height=15cm]{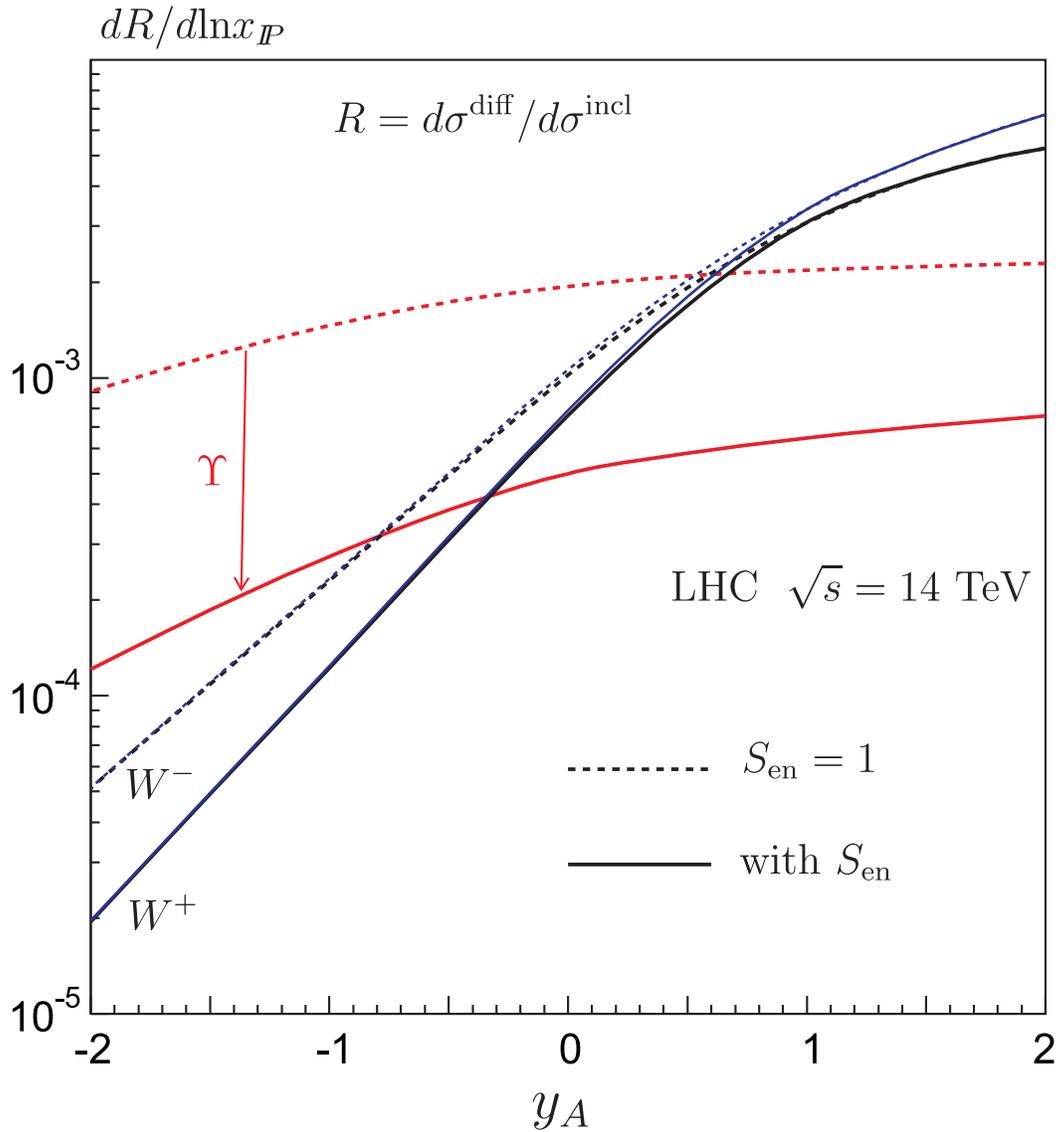}
\caption{The predictions of the ratio $R$ of (\ref{eq:Ren}) for $W$ and $\Upsilon$ production with (continuous curves) and without (dashed curves) enhanced soft rescattering on intermediate partons.}
\label{fig:upsw}
\end{center}
\end{figure}
\begin{figure} 
\begin{center}
\includegraphics[height=15cm]{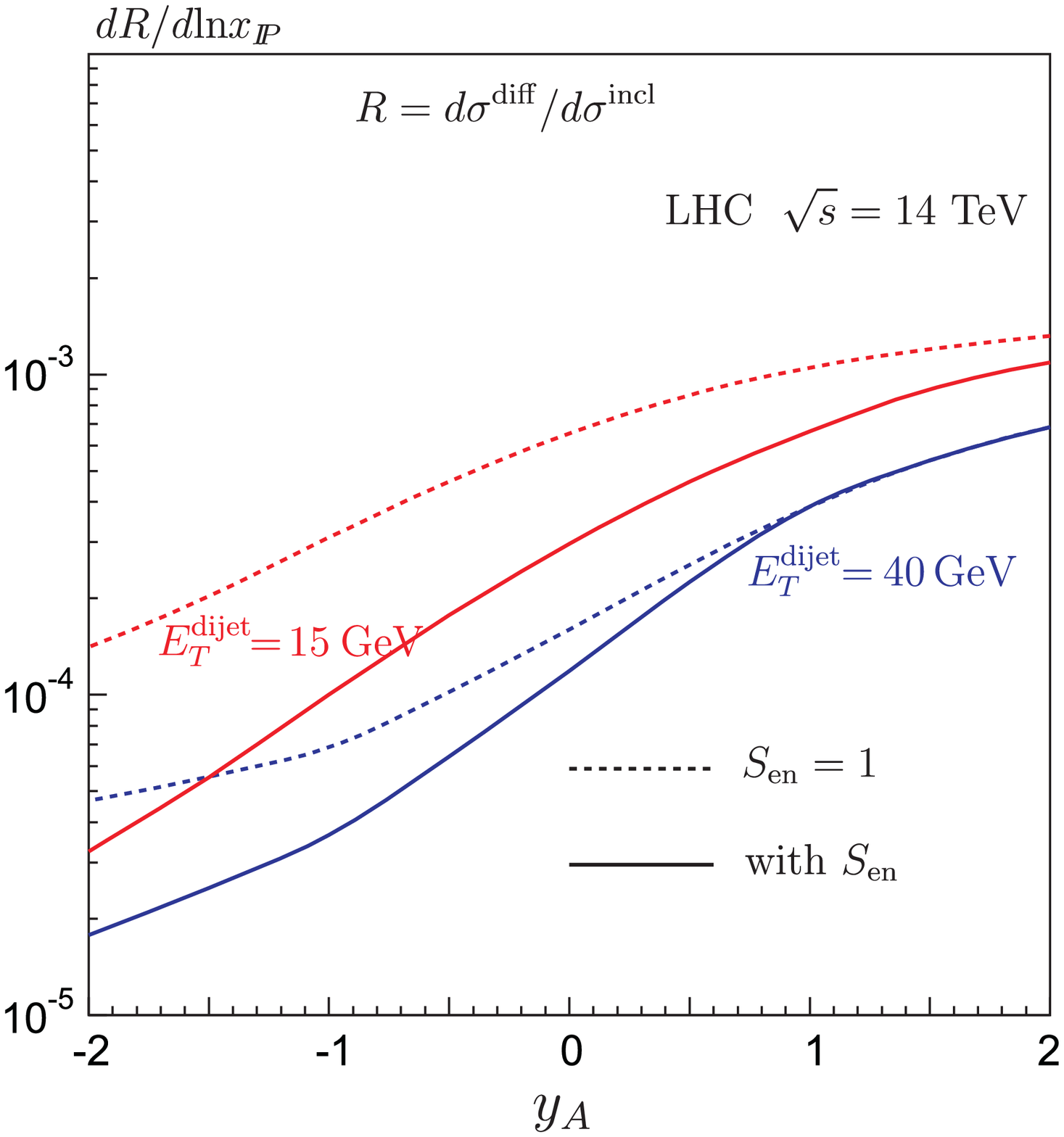}
\caption{The predictions of the ratio $R$ of (\ref{eq:Ren}) for the production of a pair of high $E_T$ jets with (continuous curves) and without (dashed curves) enhanced soft rescattering on intermediate partons.}
\label{fig:upsd}
\end{center}
\end{figure}

Experimentally, we can measure the rapidity $y_A$ of the central system ($A=W, ~\Upsilon$ or dijet) and also the momentum fraction carried in the gap direction by the soft hadrons, $\xi^- = \sum \xi_i^-$, recall the analogous sum of (\ref{eq:xi}). Thus we know the value of $x_\funp=\xi^- +\xi^-_A$. That is, we can observe a double distribution $d^2 \sigma^{\rm diff}/dx_\funp dy_A$, and form the ratio $R$ using the inclusive cross section, $d\sigma^{\rm incl}/dy_A$. If we neglect the enhanced absorption, it is straightforward to calculate the ratio $R$ of (\ref{eq:Ren}). The results are shown by the dashed curves in Figs.~\ref{fig:upsw} and \ref{fig:upsd} as a function of the rapidity $y_A$ of the heavy system $A$. When we allow for the enhanced rescattering the ratios are reduced, and lead to steeper $y_A$ distributions, as shown by the continuous curves. These plots correspond to a fixed value of $\xi^-=10~\GeV/7$ TeV. That is, we assume that the soft hadrons observed in the central calorimeter carry a longitudinal momentum\footnote{This value of $p_z$ corresponds to the realistic assumption that at the edge of the central calorimeter we will have about 2 particles with $p_t \sim$ 0.5 to 1 GeV.} $p_z$=10 GeV in the direction of the rapidity gap associated with Pomeron exchange\footnote{In this way we essentially fix the value of $x_\funp$, so that a possible interaction between the upper proton and the Pomeron placed in the gap interval (which causes an additional enhanced correction) will not affect the $y_A$ dependence of $R$.}. 

The results shown in Figs.~\ref{fig:upsw} and \ref{fig:upsd} should be regarded as an indication of the size of possible enhanced effects and not as quantitative predictions. First, the model used to estimate $S^2_{\rm en}$ is quite naive. Next, the results for large negative $y_A$ sample the diffractive gluon density at large $\beta$, where it is not well constrained. Finally, for $\Upsilon$ (and dijet production at the lower $E_T$ values) we need, for large positive $y_A$, the conventional gluon density at very small $x=\beta x_\funp \sim M_A{\rm exp}(-y_A)/\sqrt{s}\sim 10^{-4}$, where it is not well determined in the global analyses.

First, we discuss the results shown in Fig.~\ref{fig:upsw} for the ratio $R$ for $W$ production at positive $y_A$. Here, due to the large $W$ mass, we have no available rapidity interval for enhanced rescattering, that is $S^2_{\rm en}\simeq 1$. Moreover, this region corresponds to $x \sim 10^{-3}$ where the quark densities are well known. Therefore inclusive production acts as a good luminosity monitor and the ratio $R$ will yield information about the eikonal factor $S^2$. Note that we now probe the survival factor at much smaller $b_t$ than that corresponding to Figs.~\ref{fig:W2} and \ref{fig:W3}. Here we expect the eikonal (non-enhanced) survival factor to be $S^2=0.08$.

On the other hand, diffractive production of a relatively light $\Upsilon$ is associated with larger rapidity intervals available for secondaries, and hence the possibility of more soft rescattering with intermediate partons, leads to more enhanced absorption. Indeed, the expected enhanced survival factor $S^2_{\rm en}\sim 0.2-0.3$.  For $\Upsilon$ production the variation of $R$ with $y_A$ is weak. The smaller rescattering of intermediate partons with $y>y_A$ on the lower proton is compensated by stronger rescattering of the partons with $y<y_A$ on the upper proton.

Perhaps the most informative probe of $S^2_{\rm en}$ is to observe the ratio $R$ for dijet production in the region $E_T \sim 15-30$ GeV. For example for $E_T \sim$ 15 GeV we predict $S^2_{\rm en} \sim$ 0.25, 0.4 and 0.8 at $y_A=-2, ~0$ and $2$ respectively.

\section{Conclusions and Outlook}

Most of the diffractive measurements described above can be performed, without detecting the very forward protons, by taking advantage of the relatively low luminosity in the early LHC data runs. This allows the use of a veto trigger to select events with no hadronic activity in the region corresponding to the large rapidity gap(s). In this way we are able to study central exclusive diffractive processes, which should be experimentally accessible at the LHC, that probe the various individual components of the formalism used to predict their cross sections. The components are sketched in Fig.~\ref{fig:parts}.  To summarize, the gap survival factor, $S^2$, caused by eikonal rescattering may be studied as indicated in Figs.~\ref{fig:W2},~\ref{fig:W3} and \ref{fig:upsw}, and the possible enhanced, $S^2_{\rm en}$, contributions as shown in Figs.~\ref{fig:enh},~\ref{fig:upsw} and \ref{fig:upsd}. The relevant unintegrated gluon distribution, $f_g$, can be constrained by observing $\Upsilon$ production, see Fig.~\ref{fig:upsilon}, and the QCD radiative effect, $T$, may be checked by observing exclusive two- and three-jet events, see Figs.~\ref{fig:Rj}.

In the first LHC runs it may be difficult
to measure the absolute values of the cross sections with sufficient accuracy.
For instance,
it will take time to determine the
 luminosity with a precision better than, say, 10$\%$.
Thus the measurements of the ratios of the rate of events with
and without rapidity gaps (such as ($W$+gaps/$W$ inclusive) and ($Z$+gaps/$Z$ inclusive) etc.) 
will be more reliable in the early data runs.
 
When the forward proton detectors are operating, even at moderate integrated luminosity\footnote{That is, before these detectors accumulate enough luminosity to probe for new physics signals.}, much more can be done. First, it is possible to measure directly the cross section $d^2\sigma_{\rm SD}/dtdM^2_X$ for single diffractive dissociation, $pp \to p+X$, and also the cross section $d^2\sigma_{\rm DPE}/dy_1 dy_2$ for soft central diffractive production, $pp \to p+X+p$. These measurements will strongly constrain the models used to describe diffractive processes and the effects of soft rescattering. The predictions of a very recent detailed model can be found in Figs. 20 and 23 of \cite{KMRnewsoft}. It turns out that $d^2\sigma_{\rm DPE}/dy_1 dy_2$ is particularly sensitive to the detailed partonic content and sizes of the various diffractive eigenstates, see Fig. 20(d).

Next, a study of the transverse momentum distributions of both of the tagged protons, and the correlations between their momenta, $\vec{p}_{t1}$ and $\vec{p}_{t2}$, is able to scan the proton optical density (opacity) \cite{KMRtag, KMRphoton}
(see also \cite{Petrov,krp}). In principle, there are additional smearing effects caused by the intrinsic
transverse momentum spread of the proton beams, see for instance,
\cite{KP}.
Incorporation of these effects
requires detailed studies, including, in particular, the detector resolution. Therefore the predictions based on measurements of the proton
transverse momenta should allow for this smearing.

We emphasize that the selection of central exclusive dijet production in the kinematical region corresponding to the sought-after Higgs signal ($E_T \sim M_H/2$) provides an ideal ``standard candle''. At leading log accuracy, this process includes all the components of the theoretical formalism used to predict the central exclusive Higgs signal; the same parton densities in the same kinematical region, the same gap survival factors $S^2$ (and $S^2_{\rm en}$) and the same QCD radiative effect $T$. The only small difference arises at next-to-leading log order. It is caused by the interference between the large-angle soft gluon radiation from the initial active gluons and the coloured composition of the dijet system. This effect should be evaluated allowing for the detector acceptance and the selection cuts used in the experiment.

Finally note, that when the statistics allow, 
 a valuable check of 
the formalism of
central exclusive diffractive production will come from the central
diphoton production, see for instance, \cite{kmrs}.
The first CDF results \cite{cdfdiphot} are quite encouraging.

\section*{Acknowledgements}

We thank Mike Albrow, Brian Cox, Albert De Roeck, Sasha Nikitenko, Andy Pilkington,
Krzysztof Piotrzkowski and Risto Orava for encouraging us to write this article and for valuable advice.
MGR thanks the IPPP at the University of Durham  for Physics for hospitality. This work was supported by INTAS grant 05-103-7515, by grant RFBR 07-02-00023, by the Russian State grant RSGSS-5788.2006.02, and by the Russia-Israel grant 06-02-72041-204; 205; 210; 200.

\end{document}